\input harvmac
\def\half{{1 \over 2}}

\def\>{{\rangle}}
\def\<{{\langle}}

\def\p{{\partial}}

\def\L{{\Lambda}}
\def\vp{{\varphi}}

\def\a {{\alpha}}
\def\b {{\beta}}
\def\ad {{\dot a}}
\def\bd {{\dot b}}

\def\g {{\gamma}}
\def\d {{\delta}}
\def\e {{\epsilon}}

\Title{\vbox{\hbox{IFT-P.022/96}}}
{\vbox{\centerline{\bf 
Manifest Electromagnetic Duality in Closed Superstring Field Theory}}}
\bigskip\centerline{Nathan Berkovits}
\bigskip\centerline{Instituto
de F\'{\i}sica Te\'orica, Univ. Estadual Paulista}
\centerline{Rua Pamplona 145, S\~ao Paulo, SP 01405-900, BRASIL}
\bigskip\centerline{e-mail: nberkovi@snfma2.if.usp.br}
\vskip .2in
The free action for massless Ramond-Ramond fields is derived
from closed superstring field theory using the techniques of Siegel
and Zwiebach. For the uncompactified Type IIB superstring, this gives
a manifestly Lorentz-covariant action for a self-dual five-form 
field strength. 
Upon compactification to four dimensions, the action depends on  
a U(1) field strength from 4D N=2 supergravity. However, unlike the
standard Maxwell action, this action is manifestly invariant under the    
electromagnetic duality transformation which rotates $F_{mn}$ into
$\epsilon_{mnpq} F^{pq}$.  

\Date{July 1996}
%\draft
\newsec {Introduction}

Due to recent conjectures regarding non-perturbative duality symmetries
\ref\np{A. Sen, Int. J. Mod. Phys. A9 (1994) 3707\semi
C. Hull and P. Townsend, Nucl. Phys. B438 (1995) 109.}\ref\Wa
{E. Witten, Nucl. Phys. B443 (1995) 85.},
there has been an increase of interest in the Ramond-Ramond sector of
the superstring. However, problems caused by picture-changing operators
have until now prevented a string field theory analysis of the
Ramond-Ramond contribution to the action, even for the quadratic term.\ref
\zs{A. Sen and B. Zwiebach, private communication.}

This is especially obvious for the Type IIB superstring, whose massless
spectrum contains a self-dual five-form field strength for which
it is difficult to construct a manifestly
covariant action\ref\Marc
{N. Marcus and J.H. Schwarz, Phys. Lett. B115 (1982)
111.}.\foot{After breaking
manifest Lorentz covariance, an action can be constructed
for a
self-dual five-form field strength.\ref\sdual{S. Deser and
C. Teitelboim, Phys. Rev. D13 (1976) 1592\semi
M. Henneaux and C. Teitelboim, Phys. Lett. B206 (1988) 650.}\ref\SS
{A. Sen and J.H. Schwarz, Nucl. Phys. B411 (1994) 35.}
Using the methods of reference \ref\Sor{P. Pasti, 
D. Sorokin, and M. Tonin, ``Duality Symmetric Actions with Manifest
Space-Time Supersymmetry'', hep-th 9506109.},
it is possible to ``covariantize'' this action by
adding harmonic-like fields.\ref\spr{D. Sorokin,
private communication.} } Also, there are
R-R fields which allow more than one action (e.g., using
a gauge field with $P$ indices or with $D-P-2$ indices), and it is
impossible to choose the correct action without an off-shell string
field theory description.

In this paper, I begin by defining the R-R contribution to the
free action using a version of superstring
field theory which does not require picture-changing operators in either
the NS or R sectors. This version is based on the open superstring field
theory developed in reference \ref
\Siega{W. Siegel, Int. J. Mod. Phys. A6 (1991) 3997\semi
N. Berkovits, M.T. Hatsuda, and W. Siegel, Nucl. Phys. B371
(1991) 434.}, where it was shown that by
adding a non-minimal set of variables to the usual RNS variables, both
the NS and R contributions to the free 
open superstring action could be written as 
$\<\Phi |Q|\Phi\>$. Generalization to closed superstring field theory is
completely straightforward.   

I then analyze the massless R-R sector using the techniques of reference
\ref\szb{W. Siegel, Phys. Lett. B151 (1985) 396\semi
W. Siegel and B. Zwiebach, Nucl. Phys. B263 (1986)
105.} by Siegel and Zwiebach. Unlike the massless
NS-NS sector, one has an infinite
number of R-R fields due to the presence of bosonic ghost zero
modes. However, the free action for these infinite fields 
can easily be written in closed form. For the Type IIB
superstring, 
this gives a manifestly Lorentz-covariant
action for
a self-dual five-form field strength.\foot{In Hamiltonian language,
the self-duality condition involves second-class constraints. By
introducing an infinite set of fields, these second-class constraints
can be replaced with first-class constraints, allowing a covariant
expression for the path integral.\ref\Ham{
B. McClain, Y. Wu, and F. Yu, Nucl. Phys. B343 (1990) 689\semi
I. Martin and A. Restuccia, Phys. Lett. B323 (1994) 311\semi
F.P. Devecchi and M. Henneaux, ``Covariant path
integral for chiral p-forms'', hep-th 9603031.}
It would be interesting to compare
this Hamiltonian approach with the Lagrangian approach of this paper.}

After compactifying to four dimensions, one of the massless R-R fields
is the U(1) field strength $F^{mn}$ of 4D N=2 supergravity. Although    
one might expect an action
of the form $\int d^4 x F_{mn} F^{mn}$, compactification
of the ten-dimensional action gives a completely different action  
which involves infinite fields and is manifestly invariant under the
electromagnetic duality transformation 
which rotates $F_{mn}$ into
$\epsilon_{mnpq} F^{pq}$.  

In the conclusion of this paper, I discuss possible
generalizations of this work.

\newsec{Ramond-Ramond contribution to superstring field theory}

In standard open superstring field theory, the inner product of two
string fields vanishes unless the sum of the pictures of the string fields
is $-2$. This comes from the background charge on a sphere which implies
that $\< c \p c \partial^2 c~ 
e^{-2\phi}\>$ 
is non-zero. (Note that the picture $P$ of a string field $\Phi$
is defined
by $\oint dz(i\xi\eta+\partial\phi)
|\Phi\>$=$P|\Phi\>$ where $P=\oint dz(i\xi\eta+\partial\phi)$ and the
bosonic ghosts have been fermionized as $\gamma=\eta e^{\phi}$ and $\beta=
\p\xi e^{-\phi}$.\ref
\fms{D. Friedan, E. Martinec, and S. Shenker, Nucl. Phys. B271 (1986) 93.}) 

For Neveu-Schwarz string
fields, this is not a problem since the free action
$\<\Phi|Q|\Phi\>$ carries $-2$ picture if $\Phi$ is chosen to carry
$-1$ picture. However for Ramond string fields (which must carry
half-integer picture), it is clear that $\<\Phi|Q|\Phi\>$ cannot carry
$-2$ picture. For open superstring field theory, a proposed solution
was to modify the Ramond kinetic operator to $Q~Y$ where 
$Y=c\p\xi e^{-2\phi}$ is the
inverse picture-changing operator.\ref\Wft{E. Witten, Nucl. Phys. B276
(1986) 291.}
Since $Y$ carries $-1$ picture, the action
$\<\Phi|QY|\Phi\>$ carries $-2$ picture if the Ramond string
field is chosen to carry $-\half$ picture.

Although there are probably gauge-fixing problems at massive levels with
the $~$ 
$\<\Phi|QY|\Phi\>$ action
\ref\araf{I.Ya. Aref'eva and
P.B. Medvedev, Phys. Lett. B202 (1988) 510.}, 
it at least gives the correct kinetic term
for the massless Majorana-Weyl spinor, 
\eqn\act{{\cal S}=\int d^{10}x
(i s^\a \p_{\a\b} s^\b)}
where $\p_{\a\b}=\Gamma^\mu_{\a\b}\p_\mu$ and $\Gamma^\mu_{\a\b}$ is
the symmetric ten-dimensional $\sigma$-matrix.  
However, there have been no succesful generalizations of this action to
the Ramond-Ramond sector of the closed superstring.
The obvious guess, 
\eqn\guess{\<\Phi|(Q_L +Q_R)Y_L Y_R (c_{0L}-c_{0R})|\Phi\>,}
has the problem that $Y_L Y_R$ does not commute with
$b_{0L}-b_{0R}$. (Throughout this
paper, $L$ and $R$ subscripts refer to left and right-moving variables).
This means that 
the action of \guess is not invariant under the gauge transformation
$\d|\Phi\>=(b_{0L}-b_{0R})(Q_L+Q_R)|\Lambda\>$, even when $\Phi$ satisfies the
standard closed-string restrictions 
$(b_{0L}-b_{0R})|\Phi\>=
(L_{0L}-L_{0R})|\Phi\>=0$.\zs

In reference \Siega, it was shown that by
adding a new non-minimal set of variables to the usual RNS variables,
both the NS and R contributions to the free open superstring field theory
action could be written as $\<\Phi|Q|\Phi\>$. These new non-minimal variables
consist of a pair of conjugate bosons $(\tilde\g,\tilde\b)$ of weight
$(-\half,{3\over 2})$, and a pair of conjugate fermions $(\chi,u)$
of weight $(-\half,{3\over 2})$. The BRST operator is modified to
$Q_{new}=Q_{RNS}+ \int d\sigma
~u\tilde\g$, so using the standard ``quartet'' argument,
the new non-minimal fields do not affect the physical cohomology.
Like the $\psi^\mu$ matter fields and $(\g,\b)$ ghost fields, 
$(\tilde\g,\tilde\b)$ and $(\chi,u)$ are defined to be odd under
$G$-parity.

It will be convenient to fermionize $(\g,\b)$ and $(\tilde\g,\tilde\b)$ in the
following non-standard way:
\eqn\ferm{t^+=\g+i\tilde\g=\eta e^{\phi},
\quad \bar t^+=\g-i\tilde\g=\bar\eta e^{\bar\phi},}
$$t^-=\half(\b+i\tilde\b)=\p\bar\xi e^{-\bar\phi},
\quad \bar t^-=\half(\b-i\tilde\b)=\p\xi e^{-\phi},$$
where $$t^j(y)\bar t^k(z)\to \e^{jk}(y-z)^{-1},\quad\quad(\e^{+-}=-\e^{-+}=1)$$
$$\phi(y)\phi(z)\to -\log (y-z),\quad
\bar\phi(y)\bar\phi(z)\to -\log (y-z),$$
$$\eta(y)\xi(z)\to i(y-z)^{-1},\quad
\bar\eta(y)\bar\xi(z)\to i(y-z)^{-1}$$
are the singular OPE's as $y\to z$.

There are now two picture operators, 
\eqn\pict{P=\oint dz(i\xi\eta+\p\phi),\quad
\bar P=\oint dz(i\bar\xi\bar\eta+\p\bar\phi),}
which are complex conjugates of each other. One can therefore
choose the ``in'' Ramond string field, $|\Phi\>$, to carry picture
$P=-\half$ and $\bar P=-{3\over 2}$, which implies that the ``out''
Ramond string field, $\<\Phi|$, carries $\bar P=-\half$ and
$P=-{3\over 2}$. (For NS string fields, both the
``in'' and ``out'' fields carry picture $P=\bar P=-1$.)
Since the total picture for both $P$ and $\bar P$ is
$-2$, the kinetic operator no longer needs to carry picture and the
action is simply $\<\Phi|Q|\Phi\>$.
Note that the background charge now implies
that $\< c \p c\p^2 c 
~u\p u ~e^{-2(\phi+\bar\phi)}\>$ is non-zero.

As explained in reference \Siega, 
the reality conditions on $\Phi$ are slightly
unusual. Normally, one requires that hermitian conjugation is equivalent
(up to a sign) with BPZ conjugation.\ref\Bart{B. Zwiebach,
Nucl. Phys. B390 (1993) 33.}
In this case, one instead requires
that hermitian conjugation is equivalent (up to a sign) with
BPZ conjugation times ``charge conjugation'', where ``charge conjugation''
simply flips the sign of all non-minimal fields (i.e.
$[\tilde\g,\tilde\b,u,\chi]\to$
$[-\tilde\g,-\tilde\b,-u,-\chi]$). Note that charge conjugation commutes
with the BRST operator.

To generalize to the closed superstring, one simply introduces both
left-moving and right-moving non-minimal fields:
$(\tilde\g_L,\tilde\b_L,$
$u_L,\chi_L)$ and 
$(\tilde\g_R,\tilde\b_R,$
$u_R,\chi_R)$. One then defines the
Ramond-Ramond ``in'' string field, $|\Phi\>$, to have picture 
$[P_L,\bar P_L,P_R,\bar P_R]$
=$[-\half,-{3\over 2},-\half,-{3\over 2}]$, and
the Ramond-Ramond ``out'' string field, $\<\Phi|$, to have picture
$[P_L,\bar P_L,P_R,\bar P_R]$
=$[-{3\over 2},-\half,-{3\over 2},-\half]$.
As usual for closed string fields, $\Phi$ must also satisfy 
$(b_{0L}-b_{0R})|\Phi\>$=
$(L_{0L}-L_{0R})|\Phi\>=0$.\Bart 
After imposing these restrictions, the action
is simply
\eqn\simply{{\cal S}=\<\Phi|(Q_L+Q_R)(c_{0L}-c_{0R})|\Phi\>}
where $Q_L+Q_R$ 
includes the non-minimal term $\int d\sigma(\tilde\g_L u_L+\tilde\g_R u_R)$.

\newsec{Massless Ramond-Ramond contribution to the free action}

To analyze the component form of \simply, it is convenient to use the
SU(1,1) formalism developed by Siegel and Zwiebach in reference \szb.
Although this SU(1,1) method has not been generalized to include interactions,
if one only wants the quadratic term, it is much simpler than directly
performing a component expansion of \simply.
Before using this method to derive the massless Ramond-Ramond contribution
to the free closed superstring action,
I will first review 
the massless Ramond contribution to the free open superstring action.\Siega   

\subsec{The open superstring} 

For an open string action of the form $\<\Phi|Q|\Phi\>$, the first step
in the SU(1,1) method is to define $T^{++}$ and $Q^+$ by
\eqn\define{Q=c_0 L_0 +b_0 T^{++} + Q^+}
where $Q^+$ includes all terms in $Q$ which are independent of the
fermionic ghost zero modes, $b_0$
and $c_0$. The next step is to define $T^{--}$ and $T^{+-}$ in such
a way that $[T^{++},T^{+-},T^{--}]$ generate an SU(1,1) algebra.
For example, for the bosonic open string, 
\eqn\defT{T^{++}=\sum_{n>0} 2 n c_n c_{-n},\quad
T^{+-}=\half\sum_{n>0}i (c_n b_{-n}-b_n c_{-n}),\quad
T^{--}=\sum_{n>0} {1\over {2n}} b_n b_{-n}.}
(This algebra is SU(1,1) rather than SU(2) since $(T^{++})^\dagger
=T^{++}$
rather than $(T^{++})^\dagger=T^{--}$.)

Note that $2T^{+-}+i c_0 b_0$ is the ghost-number operator and
nilpotence of $Q$ implies that $[Q^+,T^{++}]=0$ and $(Q^+)^2=-T^{++}L_0$.
Therefore, $Q^+$ transforms under SU(1,1) as the top component of 
a two-component spinor $Q^j$ ($j=\pm$) which satisfies
\eqn\comm{\{Q^j, Q^k\}=-2 T^{jk} L_0.}

In reference \szb, Siegel and Zwiebach show
that any $\<\Phi|Q|\Phi\>$ action can equivalently be written as
\eqn\equival{\<\vp|c_0(2L_0 +i Q^j Q_j)|\vp\>}
where $Q_j=\e_{jk}Q^k$ and $\vp$ is an
SU(1,1) singlet which is independent of 
$c_0$ (i.e. $T^{jk}|\vp\>=b_0|\vp\>=0$).
This action has the gauge invariance
$\d|\vp\>=
Q^j |\L_j\>$ where $\L_j$ is an SU(1,1) doublet satisfying
$b_0|\L_j\>=0$. Because one only needs to consider string fields which
are SU(1,1) singlets, it is much easier to perform a component analysis
of \equival than of $\<\Phi|Q|\Phi\>$.

For the massless Ramond contribution to the free open superstring 
action, only
the zero modes in $Q$ are relevant (the positive modes all annihilate the
massless states):
\eqn\zero{Q=c_0 p^2 -b_0 \g_0^2 +\g_0\psi_0^\mu p_\mu +\tilde\g_0 u_0}
where $p_\mu=i\p_\mu$ is the momentum and 
$\{\psi_0^\mu,\psi_0^\nu\}=2\eta^{\mu\nu}$.
Before applying the SU(1,1) method, it is convenient to first perform
the unitary transformation 
\eqn\unitary{Q\to e^{\chi_0 b_0 \tilde\g_0}Q e^{-\chi_0 b_0\tilde\g_0},}
which transforms $Q$ into 
\eqn\transf{Q=(c_0+\chi_0\tilde\g_0) p^2 -b_0 (\g_0^2+\tilde\g_0^2)
 +\g_0\psi_0^\mu p_\mu +\tilde\g_0 u_0}
$$=c_0 p^2 -b_0 t^+_0 \bar t^+_0 
 +\half(t^+_0 +\bar t^+_0)\psi_0^\mu p_\mu -{i\over 2}(t_0^+ -
\bar t_0^+)( u_0 +\chi_0 p^2)$$
where $t^j$ is defined in \ferm.
For notational
convenience, I will supress the 0 subscript when it is obvious.

So using the definitions of \define,
\eqn\using{L=p^2,\quad T^{jk}=-\half(t^j \bar t^k +t^k\bar t^j),}
$$Q^j=\half(t^j+\bar t^j)\psi^\mu p_\mu-{i \over 2}(t^j-\bar t^j)
(u+\chi p^2).$$ 
Note that $[t^j,\bar t^k]=i\e^{jk}$.  

For the ``in'' string field $|\vp\>$ which has picture
$[P,\bar P]=[-\half,-{3\over 2}]$, the massless states which are
independent of $c_0$ can be written as
$$|\vp\>= f^\a (x^\mu,\psi^\mu, iu, t^+,t^-)|0_\a\>$$
where $|0_\a\>$ is an SO(9,1) Majorana-Weyl spinor whose vertex
operator in terms of the super-$SL_2$ invariant vacuum is 
\eqn\vertex{V=c e^{-\half(\phi +3\bar\phi+
i\sigma +{{i\pi}\over 2})} S_\a.}
($u=e^{-i\sigma}$, $\chi=e^{i\sigma}$, and $S_\a$
is the usual spin field of weight $5\over 8$.)
Note that $\bar t^j_0|0_\a\>=\chi_0|0_\a\>=0$.

The reality condition discussed earlier implies that 
$f^\a$ is a real function (which is the reason for the 
$e^{-{{i\pi}\over 4}}$ factor in \vertex). 
Furthermore, the GSO projection restricts $f^\a$ to have
even $G$-parity, where $\psi^\mu$, $u$, and $t^j$ all carry
odd $G$-parity. Finally, the SU(1,1) singlet condition, 
$T^{jk}|\vp\>=0$, implies that $f^\a$ is independent of $t^j$,
and therefore, 
\eqn\Ket{|\vp\>= s^\a(x)|0_\a\> +u \kappa_\b(x) |0^\b\>}
where $\psi^\mu |0_\a\>=\Gamma^\mu_{\a\b} |0^\b\>$. (There is no
factor of $i$ in \Ket since $|0^\b\>$ is fermionic, and therefore
$(u\kappa_\b |0^\b\>)^\dagger$=$-u\kappa_\b \<0^\b|$, which
goes to $u\kappa_\b |0^\b\>$ under
charge conjugation.)  

Plugging into \equival, one finds
\eqn\plug{{\cal S}=\<\vp|c_0(2L_0+i Q^j Q_j)|\vp\>}
$$=\<\vp|c(p^2 + {1\over 2} (t^j+\bar t^j)(t_j-\bar t_j)    
p_\mu \psi^\mu (u+\chi p^2))|\vp\>,$$
where $\<\vp|=\<0_\a| s^\a(x) +\<0^\b|\kappa_\b (x) u$ and
the vertex operator for $\<0_\a|$ is  
\eqn\verttwo{V=c e^{-\half(\bar\phi +3\phi+i\sigma
-{{i\pi}\over 2})} S_\a.}

Using the normalization condition that $\<0_\a| c_0 u_0 ~h(x) |0^\b\>=  $
$i\d_\a^\b \int d^{10}x h(x)$, it is easy to compute that
\eqn\openf{{\cal S}=\int d^{10}x (-2i s^\a \p_\mu \p^\mu \kappa_\a
+i s^\a \p_{\a\b} s^\b -i\kappa_\a \p^{\a\b}\p_\mu\p^\mu \kappa_\b)}
$$=\int d^{10}x (i\hat s^\a \p_{\a\b}\hat s^\b)$$
where $\hat s^\a=s^\a -\p^{\a\b}\kappa_\b$.
Note that \openf is invariant under the gauge transformation:
\eqn\noteg{\d|\vp\>= Q^j(-t_j\Lambda_\b)|0^\b\>=u\L_\b |0^\b\> + 
\p^{\a\b}\L_\b |0_\a\>,}
i.e. $\d\kappa_\b=\L_\b$, $\d s^\a=\p^{\a\b}\L_\b$.    

\subsec{The closed superstring}

For the closed superstring, the SU(1,1) method is generalized
in the obvious way. One uses $Q_L$ and $Q_R$ to define
$T^{jk}_L$, $Q^j_L$, $T^{jk}_R$, and $Q^j_R$, and the action
is
\eqn\clos{\<\vp|c_{0L} c_{0R}(2L_{0L}
+i (Q^j_L+Q^j_R)(Q_{jL}+Q_{jR}))|\vp\>}
where $|\vp\>$ satisfies 
\eqn\rest{
b_{0L}|\vp\>=b_{0R}|\vp\>=(L_{0L}-L_{0R})|\vp\>=(T^{jk}_L+T^{jk}_R)|\vp\>
=0.}
This action has the gauge invariance
\eqn\clg{\d|\vp\>=(Q^j_L+Q^j_R)|\L_j\>}
where $b_{0L}|\L_j\>$ 
= $b_{0R}|\L_j\>=0$ and  
$\L_j$ transforms as an SU(1,1) doublet under
$T^{jk}_L+T^{jk}_R$.  
 
For the ``in'' closed-string field with picture
$[P_L,\bar P_L,$
$P_R,\bar P_R]$=
$[-\half,-{3\over 2},$
$-\half,-{3\over 2}]$,
the massless states of the Type IIB superstring can be written as
\eqn\massl{|\vp\>= f^{\a\b} (x^\mu,\psi^\mu_{L},\psi^\mu_{R}, iu_L,
iu_R, t^j_L,t^j_R)|L_{\a}\>|R_\b\> }
where $|L_\a\>$ and $|R_\b\>$ are 
defined like $|0_\a\>$ in the open superstring
($|L_\a\>$ is left-moving
and $|R_\b\>$ is right-moving). $f^{\a\b}$ is real and, for the Type IIB
superstring, has 
even left and right-moving $G$-parity. (For the Type IIA superstring,
$f^{\a\b}$ has even left-moving and odd right-moving $G$-parity).  
Furthermore, the SU(1,1) singlet condition,  
$(T^{jk}_L+T^{jk}_R)|\vp\>=0$, implies that $f^{\a\b}$ depends on
$t^j_L$ and $t^j_R$ only in the combination $\e_{jk} t^j_L t^k_R$.
Therefore, for Type IIB, 
$$|\vp\>=\sum_{n=0}^\infty {1\over {(2n+1)!}}
(it^j_L t_{jR})^{2n} 
(F_{(2n)}^{\a\b}
|L_\a\> |R_\b\>  +u_L E_{(2n)\a}^\b 
|L^\a\> |R_\b\> $$
\eqn\closex{ +u_R D_{(2n)\b}^\a 
|L_\a\> |R^\b\>  +u_L u_R  C_{(2n)\a\b}  
|L^\a\> |R^\b\> )}
$$+\sum_{n=0}^\infty {1\over{(2n+2)!}}
(it^j_L t_{jR})^{2n+1}
(F_{(2n+1)\a\b}
|L^\a\> |R^\b\>  +u_L E_{(2n+1)\b}^\a
|L_\a\> |R^\b\> $$
$$ +u_R D_{(2n+1)\a}^\b 
|L^\a\> |R_\b\>  +u_L u_R  C_{(2n+1)}^{\a\b}  
|L_\a\> |R_\b\> ).$$

So at the massless level, an infinite number of fields are present
in the Ramond-Ramond sector of the closed 
superstring. (For the Type IIA superstring, the
only difference is in the position of the right-moving spinor
index on $F$, $E$, $D$, and $C$.) 

Plugging into \clos and keeping only the zero modes, one derives that
the massless Ramond-Ramond contribution to the free action is:
\eqn\derivec{{\cal S}= 
\<\vp|c_{0L} c_{0R}(2L_{0L}
+i \e_{jk} (Q^j_L+Q^j_R)(Q^k_L+Q^k_R))|\vp\>}
$$=\<\vp|c_L c_R(
(i-t_L^j \bar t_{jL} )p_\mu\psi_L^\mu (u_L+\chi_L p^2)  
+(i-t_R^j \bar t_{jR})p_\mu\psi_R^\mu (u_R+\chi_R p^2) $$
$$+{1\over 2}
((t_L^j+\bar t_L^j)(t_{jR}-\bar t_{jR}) p_\mu\psi_L^\mu (u_R+\chi_R p^2) 
+
(t_L^j-\bar t_L^j)(t_{jR}+\bar t_{jR})(u_L+\chi_L p^2) p_\mu\psi_R^\mu)$$ 
$$+{i\over 2}
(t_L^j+\bar t_L^j)(t_{jR}+\bar t_{jR})p_\mu\psi_L^\mu  p_\nu\psi_R^\nu
$$
$$-{i\over 2}
(t_L^j-\bar t_L^j)(t_{jR}-\bar t_{jR})
(u_L+\chi_L p^2)(u_R+\chi_R p^2))|\vp\>.$$

Finally, using the fact that 
$$\<R_\a| \<L_\b| (\bar t^j_L\bar t_{jR})^m 
(t^j_L t_{jR})^n 
~c_{0L} c_{0R}~ u_{0L} u_{0R}~ h(x)|L^\g\> |R^\d\>$$
\eqn\normc{= \d_{n,m}~ n! (n+1)! ~(-1)^n 
\d_\a^\d \d_\b^\g \int d^{10}x h(x),}
it is straightforward to compute that 
$${\cal S}=  
\sum_{n=0}^\infty [
2C_{(n)\a\b}\p_\mu\p^\mu(\p^{\a\g}E_{(n)\g}^\b +
(-1)^n\p^{\b\g}D^\a_{(n)\g})$$ 
$$-2F_{(n)}^{\a\b}((-1)^n\p_{\b\g}E_{(n)\a}^\g +\p_{\a\g}D^\g_{(n)\b}) $$
\eqn\finalc{
-(F^{\a\b}_{(n)}+\p^{\a\g}E_{(n)\g}^\b+(-1)^n\p^{\b\g}D_{(n)\g}^\a
-(-1)^n\p^{\a\g}\p^{\b\d}C_{(n)\g\d})}
$$(F_{(n+1)\a\b}+(-1)^n\p_{\b\kappa}D_{(n+1)\a}^\kappa
-\p_{\a\kappa}E^\kappa_{(n+1)\b}
+(-1)^n\p_{\a\kappa}\p_{\b\e}C_{(n+1)}^{\kappa\e})]$$
where it is understood that for odd $n$, the positions of all spinor 
indices are reversed (e.g. the first term for $n=1$ is
$2C_{(1)}^{\a\b} \p_\mu\p^\mu\p_{\a\g}E_{(1)\b}^\g$).

Although \finalc looks complicated, it is easy to analyze because
of the gauge invariance: 
\eqn\easy{\d|\vp\>= Q^j\sum_{n=0}^\infty {1\over
{(n+2)!}}(it_L^k t_{Rk})^{n} }
$$
(-2t_{jL}\L_{(n)\a}^\b|L^\a\> |R_\b\> 
-2t_{jR}\Omega_{(n)\b}^\a|L_\a\> |R^\b\> 
-2t_{jL}u_R\Theta_{(n)\a\b}|L^\a\> |R^\b\>)$$
$$=\sum_{n=0}^\infty {{(it_L^j t_{jR})^n}\over{(n+1)!}}
(u_L u_R\Theta_{(n)\a\b} |L^\a\>|R^\b\>
+u_R(\Omega_{(n)\b}^\a-\p^{\a\g}\Theta_{(n)\g\b}) |L_\a\>|R^\b\>$$
$$+ u_L\L_{(n)\a}^\b |L^\a\>|R_\b\>+
(\p^{\a\g}\L_{(n)\g}^\b+(-1)^n\p^{\b\g}\Omega_{(n)\g}^\a )|L_\a\>|R_\b\>)
$$
$$
+\sum_{n=0}^\infty {{(it_L^j t_{jR})^{n+1}}\over{(n+2)!}}
(u_R(-\L_{(n)\a}^\b +(-1)^n\p^{\b\g}\Theta_{(n)\a\g})|L^\a\>|R_\b\>+
u_L\Omega_{(n)\b}^\a |L_\a\>|R^\b\>+
$$
$$(-(-1)^n\p_{\b\g}\L_{(n)\a}^\g-\p_{\a\g}\Omega_{(n)\b}^\g+
\p_\mu\p^\mu\Theta_{(n)\a\b} )|L^\a\>|R^\b\>)$$
where $\L_{(n)\a}^\b$, $\Omega_{(n)\b}^\a$, and $\Theta_{(n)\a\b}$ are 
independent fields (for odd $n$, the positions of their spinor indices
are reversed).

Comparing with \closex, this implies that \finalc is gauge-invariant under
\eqn\comp{\d C_{(n)\a\b}=\Theta_{(n)\a\b},\quad 
\d D_{(n)\b}^\a=\Omega_{(n)\b}^\a-\p^{\a\g}\Theta_{(n)\g\b},}
$$\d E_{(n)\a}^\b=\L_{(n)\a}^\b,\quad
\d F_{(n)}^{\a\b}=\p^{\a\g}\L_{(n)\g}^\b+
(-1)^n\p^{\b\g}\Omega_{(n)\g}^\a,$$
$$\d D_{(n+1)\a}^\b= 
-\L_{(n)\a}^\b +(-1)^n\p^{\b\g}\Theta_{(n)\a\g},\quad
\d E_{(n+1)\b}^\a=\Omega_{(n)\b}^\a,$$
$$
\d F_{(n+1)\a\b}=
-(-1)^n\p_{\b\g}\L_{(n)\a}^\g-\p_{\a\g}\Omega_{(n)\b}^\g+
\p_\mu\p^\mu\Theta_{(n)\a\b}.$$ 

Therefore, 
$\L_{(n)\a}^\b$, $\Omega_{(n)\b}^\a$, and $\Theta_{(n)\a\b}$ can
be used to gauge
\eqn\gef{E_{(n)\a}^\b=D_{(n)\b}^\a=C_{(n)\a\b}=0}
for all $n$. In this gauge,
$$|\vp\>=\sum_{n=0}^\infty ({{(it_L^j t_{jR})^{2n}}\over
{(2n+1)!}} F_{(2n)}^{\a\b}|L_\a\> |R_\b\>+
{{(it_L^j t_{jR})^{2n+1}}\over{(2n+2)!}} F_{(2n+1)\a\b}|L^\a\> |R^\b\> ),$$
and the equation of motion
\eqn\motion{(2L_{0L} +i(Q^j_L+Q^j_R)(Q_{jL}+Q_{jR}))|\vp\>=0}
implies that
\eqn\one{ 0=F_{(1)\a\b}=-F_{(3)\a\b}=F_{(5)\a\b}=-F_{(7)\a\b}= ... ,}
\eqn\two{ 
F_{(0)}^{\a\b}=-F_{(2)}^{\a\b}=F_{(4)}^{\a\b}=-F_{(6)}^{\a\b}= ... ,}
\eqn\three{ \p_{\a\g}F_{(0)}^{\a\b}=
\p_{\b\g}F_{(0)}^{\a\b}=0.}
Equation \one comes from the 
$u_L u_R$
$ (t_L^j t_{jR})^{2n}$ component of \motion,
equation \two comes from the $u_L u_R$
$ (t_L^j t_{jR})^{2n+1}$ component of \motion,
and equation \three comes from the $u_L$ and $u_R$ components of \motion.

So, on-shell, the only non-vanishing field in this gauge is $F_{(0)}^{\a\b}$,
which satisfies the equations of motion of \three.
Expanding in vector notation,
\eqn\vect{F_{(0)}^{\a\b}=\Gamma_\mu^{\a\b} F^\mu_{(0)}+
\Gamma_{\mu\nu\rho}^{\a\b} F^{\mu\nu\rho}_{(0)}+
\Gamma_{\mu\nu\rho\kappa\sigma}^{\a\b} F^{\mu\nu\rho\kappa\sigma}_{(0)}}
where 
$F^{\mu\nu\rho\kappa\sigma}_{(0)}$ is self-dual. Equation \three implies that
these one-form, three-form, and self-dual five-form field strengths obey 
\eqn\bianch{\p^{[\nu} F^{\mu]}_{(0)}=
\p^{[\nu} F^{\mu\rho\kappa]}_{(0)}=0}
\eqn\mot{\p_\mu F^{\mu}_{(0)}=
\p_\mu F^{\mu\nu\rho}_{(0)}=
\p_\mu F^{\mu\nu\rho\kappa\sigma}_{(0)}=0,}
which are the usual Bianchi identities and equations of motion.
(For the Type IIA superstring, one gets Bianchi identities and
equations of motion for a two-form and a four-form field strength.)

It might seem puzzling that in the gauge of \gef, the action of \finalc
becomes
\eqn\puz{{\cal S}=-\sum_{n=0}^\infty F_{(n)}^{\a\b} F_{(n+1)\a\b},}
which naively appears to have no propagating degrees of freedom. This 
paradox comes from the fact that there are an infinite number of fields
and one needs to be careful when taking the limit $n\to\infty$.

The correct procedure is to first compute the equations of motion with
a finite cutoff, so $n$ ranges from 0 to $N$. If $N$ is finite, one
can gauge away [$E_{(n)}$,$D_{(n)}$,$C_{(n)}$] for $n<N$,
but cannot gauge away 
[$E_{(N)}$,$D_{(N)}$,$C_{(N)}$] since the gauge transformations of \comp 
parameterized by 
[$\L_{(N)}$,$\Omega_{(N)}$,$\Theta_{(N)}$] also transform
[$F_{(N+1)}$,$E_{(N+1)}$,$D_{(N+1)}$,$C_{(N+1)}$].
In this gauge, one gets the equations of motion of \one\two\three
for $n<N$, as well
as non-trivial equations of motion for
[$F_{(N)}$,$E_{(N)}$,$D_{(N)}$,$C_{(N)}$].
In the limit as $N\to\infty$, one can ignore
[$F_{(N)}$,$E_{(N)}$,$D_{(N)}$,$C_{(N)}$], so one is
left with just the equations of motion of \one\two\three.

\newsec{Manifest electromagnetic duality in four dimensions}

After compactifying the Type II superstring on a generic six-dimensional
Calabi-Yau manifold, the resulting superstring contains at least N=2
four-dimensional supersymmetry. For each massless ten-dimensional 
Ramond-Ramond
field $F^{\a\b}$
with $16\times 16$ components, there always remains at least one massless
four-dimensional Ramond-Ramond field with $4\times 4$ components. Using
two-component Weyl notation, this massless four-dimensional field splits into
$F_{ab}$ and $\tilde F_{a\dot b}$, and their complex conjugates, 
$\bar F_{\dot a\dot b}$ and $\bar{\tilde F}_{\dot a b}$ ($a,b$=1 or 2).

So from the analysis of the previous section, the massless four-dimensional
Ramond-Ramond fields always contain the following fields for $n=0$ to 
$\infty$: 
\eqn\fourc
{F_{(n) a b},E_{(n)\dot a b},D_{(n) a\dot b},C_{(n)\dot a\dot b},\quad
\bar F_{(n)\dot a \dot b},\bar E_{(n) a\dot b},\bar D_{(n)\dot a b},
\bar C_{(n) a b},}
$$\tilde F_{(n) a\dot b},
\tilde E_{(n)\dot a \dot b},\tilde D_{(n) a b},\tilde C_{(n)\dot a b},\quad
\bar{\tilde F}_{(n) \dot a b},
\bar{\tilde E}_{(n) a b},
\bar{\tilde D}_{(n)\dot a\dot b},\bar{\tilde C}_{(n) a\dot b}.$$

The free action for these four-dimensional massless fields is easily 
obtained from \finalc
by ignoring 12 of the 16 components of each ten-
dimensional
spinor. The resulting action is:
$${\cal S}=  
\sum_{n=0}^\infty [
2C_{(n)\ad\bd}\p_\mu\p^\mu(\p^{a\ad}
\bar E_{(n)a}^\bd +(-1)^n\p^{b\bd}\bar D^\ad_{(n)b}) $$
$$-2F_{(n)}^{ab}((-1)^n\p_{b\bd}\bar E_{(n)a}^\bd +\p_{a\ad}
\bar D^\ad_{(n)b})$$ 
\eqn\actionfour{
-(F^{ab}_{(n)}+\p^{a\ad}E_{(n)\ad}^b+(-1)^n\p^{b\bd}D_{(n)\bd}^a
-(-1)^n\p^{a\ad}\p^{b\bd}C_{(n)\ad\bd})}
$$(F_{(n+1)ab}+(-1)^n\p_{b\dot c}D_{(n+1)a}^{\dot c}
-\p_{a\dot c}E^{\dot c}_{(n+1)b}
+(-1)^n\p_{a\dot c}\p_{b\dot d}C_{(n+1)}^{\dot c\dot d})]$$

$$+\sum_{n=0}^\infty [
2\tilde C_{(n)\ad b}\p_\mu\p^\mu(\p^{a\ad}
\bar{\tilde E}_{(n)a}^b +(-1)^n\p^{b\bd}\bar{\tilde D}^\ad_{(n)\bd}) $$
$$-2\tilde F_{(n)}^{a\bd}((-1)^n\p_{b\bd}\bar{\tilde E}_{(n)a}^b +\p_{a\ad}
\bar{\tilde D}^\ad_{(n)\bd}) $$ 
$$-(\tilde F^{a\bd}_{(n)}+\p^{a\ad}
\tilde E_{(n)\ad}^\bd+(-1)^n\p^{b\bd}\tilde D_{(n)b}^a
-(-1)^n\p^{a\ad}\p^{b\bd}\tilde C_{(n)\ad b})$$
$$(\tilde F_{(n+1)a\bd}+(-1)^n\p_{c\bd}\tilde D_{(n+1)a}^{c}
-\p_{a\dot c}\tilde E^{\dot c}_{(n+1)\bd}
+(-1)^n\p_{a\dot c}\p_{d\dot b}\tilde C_{(n+1)}^{\dot c d})]$$
$$+~complex~conjugate.$$
Note that tilded and un-tilded fields do not couple in \actionfour, 
which is
related to the non-coupling of 
hypermultiplets and vector multiplets in low-energy N=2 actions.

After gauge-fixing 
\eqn\gfour{E_{(n)}^{\ad b}=D_{(n)}^{a\bd}
=C_{(n)}^{\ad\bd}=
\tilde E_{(n)}^{\ad\bd}=\tilde D_{(n)}^{ab}=\tilde C_{(n)}^{\ad b}
=0}
as in \gef, the equations of motion are
\eqn\fone{ 0=F_{(1)}^{ab}=-F_{(3)}^{ab}=F_{(5)}^{ab}= ... ,\quad
0=\tilde F_{(1)}^{a\bd}=-\tilde F_{(3)}^{a\bd}
=\tilde F_{(5)}^{a\bd}= ... ,}
\eqn\ftwo{ F_{(0)}^{ab}=-F_{(2)}^{ab}=F_{(4)}^{ab}= ... ,\quad 
\tilde F_{(0)}^{a\bd}
=-\tilde F_{(2)}^{a\bd}=\tilde F_{(4)}^{a\bd}= ... ,}
\eqn\fthree{ \p_{a\ad}F_{(0)}^{ab}=
\p_{b\bd}F_{(0)}^{ab}=0,\quad
 \p_{a\ad}\tilde F_{(0)}^{a\bd}=
\p_{b\bd} \tilde F_{(0)}^{a\bd}=0
.}

Using the vector notation,
\eqn\vecf
{F_{(0)}^{ab}=F_{(0)} \e^{ab} + F_{(0)}^{mn} \sigma_{mn}^{ab} ,\quad
\tilde F_{(0)}^{a\bd} =\tilde F^m_{(0)} \sigma_m^{a\bd},}
one gets the equations of motion from \fthree:
\eqn\get{\p_m F_{(0)} =\p_m F^{mn}_{(0)}=\p_m\tilde F^m_{(0)}=0,}
\eqn\fbianch{\e_{mnpq}\p^m F^{np}_{(0)}=\e_{mnpq}\p^m\tilde F^n_{(0)}=0.}
(Note that 
$F_{(0)}^{mn}$ is real, while $F_{(0)}$ and
$\tilde F^m_{(0)}$ are complex.)

Equations \fbianch imply that $F^{mn}_{(0)}$ and $\tilde F^m_{(0)}$ can be 
expressed in terms of a real U(1) gauge field $A_m$ and a complex scalar
$y$ as
$$F^{mn}_{(0)}=\p^m A^n-\p^n A^m,\quad\tilde F^m_{(0)}=\p^m y.$$
Equations 
\get imply that $A_m$ and $y$ propagate on-shell and that $F_{(0)}$ is
a constant. Together with the NS-NS graviton, axion, and dilaton, the R-R
fields $A_m$, $y$, and $\bar y$ form the bosonic degrees of freedom of
an N=2 supergravity and dilaton multiplet.    

Although all previous descriptions 
of the four-dimensional R-R U(1) gauge field
assumed an action of the form $\int d^4 x F^{mn} F_{mn}$, it is seen from 
\actionfour
that the action coming from closed superstring field theory is completely
different 
and requires an infinite number of fields. As will now be shown, the
action of \actionfour
is manifestly invariant under the electromagnetic duality
transformation, $F^{mn}\to\e^{mnpq} F_{pq}$, whereas for the usual 
$\int d^4 x F^{mn} F_{mn}$ action, only the equations of motion are invariant.

In spinor notation, the electromagnetic duality transformation is
\eqn\spin{F^{ab}\to iF^{ab},\quad \bar F^{\ad\bd}\to -i\bar F^{\ad\bd},}
or in its continuous version,
\eqn\cont{F^{ab}\to e^{i\theta} F^{ab},\quad \bar F^{\ad\bd}\to e^{-i\theta}
\bar F^{\ad\bd}}
where $\theta$ is an arbitrary real constant.

It is easy to see that the action of 
\actionfour is invariant under
$F_{(0)}^{ab}\to e^{i\theta}F_{(0)}^{ab}$ if the un-tilded
fields
transform as
\eqn\other{
F_{(2n)}^{ab}\to e^{i\theta}F_{(2n)}^{ab},\quad
E_{(2n)}^{\ad b}\to e^{i\theta}E_{(2n)}^{\ad b},}
$$D_{(2n)}^{a \bd}\to e^{i\theta}D_{(2n)}^{a\bd },\quad
C_{(2n)}^{\ad \bd}\to e^{i\theta}C_{(2n)}^{\ad\bd },$$
$$
F_{(2n+1)}^{ab}\to e^{-i\theta}F_{(2n+1)}^{ab},\quad
E_{(2n+1)}^{\ad b}\to e^{-i\theta}E_{(2n+1)}^{\ad b},$$
$$
D_{(2n+1)}^{a \bd}\to e^{-i\theta}D_{(2n+1)}^{a\bd },
C_{(2n+1)}^{\ad \bd}\to e^{-i\theta}C_{(2n+1)}^{\ad\bd },$$
and the tilded fields are left unchanged.
So, unlike the 
$\int d^4 x F^{mn} F_{mn}$ action, the 
action of \actionfour is manifestly 
invariant under the electromagnetic duality
transformation of \spin.

\newsec{Conclusions}

In this paper, I have derived the massless Ramond-Ramond contribution to the
free action using closed superstring field theory.
This action contains an infinite number of fields and, after
compactification to four dimensions, is manifestly invariant under
electromagnetic duality transformations.

There are various possible generalizations of this work. One
generalization would be to combine the NS-NS, NS-R, R-NS, and R-R
contributions into a manifestly spacetime-supersymmetric action.
After compactification to four dimensions, this is already
possible using a closed superstring field theory\ref\prog
{N. Berkovits, work in progress.}
based on the new spacetime-supersymmetric description of the
superstring\ref\me{N. Berkovits, ``A New Description of the
Superstring'', to appear in proceedings of the
VIII J.A. Swieca summer school, hep-th 9604123\semi
N. Berkovits, Nucl. Phys. B459 (1996) 439.}.
(In fact, this was how the action of \actionfour was
originally discovered.)

A second generalization would be to construct the complete
interacting contribution of the R-R fields. It would be very
interesting to see if manifest electromagnetic duality is preserved
at the interacting level. It would also be interesting to compute
the dilaton coupling of R-R fields and see if standard arguments\Wa
based on $\int d^4 x F^{mn}F_{mn}$-type actions need to be revised.

A third generalization would be to extend the manifest electromagnetic
duality to SL(2,R) duality by including interactions with scalar fields.
As discussed in reference \SS, SL(2,R) duality is naturally manifest
in an action coming from five-branes, as opposed to T-duality, which is
naturally manifest in an action coming from strings. However, in this case,
both T-duality and SL(2,R) duality would be manifest in an action coming
from strings.

\vskip 20pt 

{\bf Acknowledgements:} I would like to thank Ashok Das, Martin Ro\^cek,
Warren Siegel, and Barton Zwiebach for useful conversations. This
work was financially supported by the Brazilian
Conselho Nacional de Pesquisa.

\listrefs
\end